\documentclass[%
aip,
reprint,
superscriptaddress,
ymb,
prb,
twocolumn,
titlepage,
floatfix,
showpacs,
]{revtex4-2}

\usepackage{graphicx}
\usepackage{hyperref}
\usepackage{amssymb}
\usepackage[fleqn]{amsmath}
\usepackage{textcomp}
\usepackage{xcolor}
\usepackage{ulem}
\allowdisplaybreaks 

\definecolor{linkblue}{RGB}{49,49,148}
\hypersetup{linkcolor  = linkblue,citecolor  = linkblue,urlcolor   = linkblue,colorlinks = true,}

\newcommand*{\bracite}[1]{%
  \textsuperscript{[}\cite{#1}\textsuperscript{]}}

\makeatletter
  \renewcommand*{\fnum@figure}{{\sffamily\bfseries\figurename~\thefigure}}
  \renewcommand*{\@caption@fignum@sep}{\textbf{. }\sffamily}
  \renewcommand{\figurename}{Figure}
  \renewcommand*{\fnum@table}{{\sffamily\bfseries\tablename~\thetable}}
  \renewcommand*{\@caption@fignum@sep}{\textbf{. }\sffamily}
  \renewcommand{\tablename}{Table}
  \renewcommand{\thetable}{\arabic{table}} 
  
\makeatother

\makeatletter
\renewcommand*{\eqref}[1]{%
  \hyperref[{#1}]{\textup{\tagform@{\ref*{#1}}}}%
}
\makeatother

\begin{document}

\title{Controlling effective field contributions to laser-induced magnetization precession by heterostructure design}

\author{J.~Jarecki}
\affiliation{Institut f\"ur Physik \& Astronomie,  Universit\"at Potsdam,  14476 Potsdam,  Germany}
\affiliation{Max-Born-Institut f\"ur Nichtlineare Optik und Kurzzeitspektroskopie,  12489 Berlin,  Germany}

\author{M.~Mattern}
\email{mamattern@uni-potsdam.de}
\affiliation{Institut f\"ur Physik \& Astronomie,  Universit\"at Potsdam, 14476 Potsdam, Germany}

\author{F.-C.~Weber}
\affiliation{Institut f\"ur Physik \& Astronomie,  Universit\"at Potsdam, 14476 Potsdam, Germany}

\author{J.-E.~Pudell}
\affiliation{Institut f\"ur Physik \& Astronomie,  Universit\"at Potsdam, 14476 Potsdam, Germany}
\affiliation{European X-Ray Free-Electron Laser Facility, Schenefeld, Germany}
\affiliation{Helmholtz-Zentrum Berlin f\"ur Materialien und Energie, BESSY~II, 12489 Berlin, Germany}

\author{X.-G. Wang}
\affiliation{School of Physics and Electronics, Central South University, Changsha 410083, China}

\author{J.-C.~Rojas S\'{a}nchez}
\affiliation{Institut Jean Lamour (UMR CNRS 7198), Universit\'e Lorraine,  54000 Nancy,   France}

\author{M.~Hehn}
\affiliation{Institut Jean Lamour (UMR CNRS 7198), Universit\'e Lorraine,  54000 Nancy,   France}

\author{A.~von~Reppert}
\affiliation{Institut f\"ur Physik \& Astronomie,  Universit\"at Potsdam,  14476 Potsdam, Germany}

\author{M.~Bargheer}
\email{bargheer@uni-potsdam.de}
\affiliation{Institut  f\"ur Physik \& Astronomie, Universit\"at Potsdam, 14476 Potsdam, Germany}
\affiliation{Helmholtz-Zentrum Berlin f\"ur Materialien und Energie, BESSY~II, 12489 Berlin, Germany}

\date{\today}

\begin{abstract}

Nanoscale heterostructure design can control laser-induced heat dissipation and strain propagation as well as their efficiency for driving magnetization precession. We use insulating MgO layers incorporated into metallic Pt-Cu-Ni heterostructures to block the propagation of hot electrons. Ultrafast x-ray diffraction (UXRD) experiments quantify how this enables controlling the spatio-temporal shape of the transient heat and strain, which drive the magnetization dynamics in the Ni layer. The frequency of the magnetization precession observed by the time-resolved magneto-optical Kerr effect (MOKE) in polar geometry is systematically tuned by the magnetic field orientation. The combined experimental analysis (UXRD and MOKE) and modeling of transient strain, heat and magnetization uniquely highlights the importance of quasi-static strain as a driver of precession, when the magnetic material is rapidly heated via electrons. The concomitant effective field change originating from demagnetization partially compensates the change induced by quasi-static strain. Tailored strain pulses shaped via the nanoscale heterostructure design provide an equally efficient, phase-matched driver of precession, paving the way for opto-magneto-acoustic devices with low heat energy deposited in the magnetic layer.
\end{abstract}

%Nanoscale heterostructure design can control laser-induced heat dissipation and strain propagation as well as their efficiency for driving magnetization precession. Insulating MgO layers incorporated into metallic Pt-Cu-Ni heterostructures block the propagation of hot electrons. Ultrafast x-ray diffraction (UXRD) experiments quantify how this enables controlling the spatio-temporal shape of the transient heat and strain, which drive the magnetization dynamics in the Ni layer. The frequency of the magnetization precession observed by the time-resolved magneto-optical Kerr effect (MOKE) in polar geometry is systematically tuned by the magnetic field orientation. The combined experimental analysis (UXRD and MOKE) and modeling of transient strain, heat and magnetization uniquely highlights the importance of quasi-static strain as a driver of precession. It can prevail over the demagnetization-induced precession when the magnetic material is rapidly heated via electrons. Tailored strain pulses shaped via the nanoscale heterostructure design provide an efficient, phase-matched driver of precession, paving the way for opto-magneto-acoustic devices.

\maketitle

\section{\label{sec:I}Introduction} 
Direct optical excitation of ferromagnets triggers phenomena such as ultrafast demagnetization\bracite{beau1996, hohl1997, stam2010, koop2010, you2018}, coherent magnetization precession\bracite{ma2015, bigo2005, kamp2002b, kats2016, ma2015, kamp2002a, shin2022, shin2023, sche2010, kim2012, jage2013, bomb2013, kim2015, kim2017}, spin-transport\bracite{razd2017, igar2023, rong2023} and all-optical magnetization switching\bracite{aleb2012, mang2014, stan2007}. An optimized excitation of coherent magnetization precession enables precessional switching\bracite{kova2013, band2021, yang2021, dolg2023}. Most investigations have focused on mechanisms using direct excitation of the magnetically ordered material, to drive precession via magnetic anisotropy changes or via magneto-elastic coupling:  magnetocrystalline anisotropy is controlled by temperature\bracite{ma2015, bigo2005, kamp2002b, kats2016}, shape anisotropy by  demagnetization\bracite{ma2015, kamp2002a, shin2023}, and strain effects are distinguished as resulting from quasi-static expansion\bracite{shin2022, shin2023} and propagating strain pulses\bracite{sche2010, kim2012, jage2013, bomb2013, kim2015, kim2017,yang2021, vern2022}. 

In a single ferromagnetic layer, the optically deposited energy leads to all at once: Ultrafast demagnetization, temperature-induced anisotropy changes and both quasi-static and propagating strain components, which are all a consequence of the rapid rise of electron-, spin- and phonon-temperatures. Each of these mechanisms can be described as pump-induced changes of effective field contributions that drive coherent magnetization precession, although the effects can cancel each other - calling for handles that turn individual effects on and off. Analysis and control of the induced precession has been mainly limited to the external magnetic field\bracite{jage2013, ma2015}, the layer thickness determining the round trip time of acoustic strain pulses\bracite{kim2012, kim2017, vern2022} or multipulse excitation schemes tuning the strain pulse pattern\bracite{kim2017, boja2015}. 

Integrating the functional magnetic layer into nanoscale heterostructures adds multiple control scenarios: The heat transport and the propagating picosecond strain pulses can be tuned individually by transparent capping layers\bracite{lee2005, zeus2019}, metal-insulator superlattices\bracite{trig2008, jage2015} and functional transducers with tunable stress generation mechanisms via temperature and fluence\bracite{matt2021, repp2020, pude2019} or magnetic fields\bracite{matt2023a}. In addition, the combination of metals with different electron-phonon coupling strength and insulating interlayers modify the heat transport within the heterostructure\bracite{pude2018, herz2022, matt2022, matt2023b, pude2020, fert2017, fert2019}, which enables the control of both the launched propagating strain pulses and the excitation of the functional magnetic layer.

Recently, metallic heterostructures built of an optically excited transducer, a propagation layer with weak electron-phonon coupling and a buried functional magnetic layer have become intensively investigated\bracite{pude2020, deb2018, deb2021b, zeus2022, xu2017, igar2023, fert2017, fert2019, berg2016, berg2020, razd2017}. Laser-excited hot electrons travelling through the propagation layer induce ultrafast demagnetization in buried metallic magnetic layers\bracite{berg2016, berg2020, fert2017, fert2019}, drive magnetization switching\bracite{xu2017, igar2023} and excite precession\bracite{deb2018} and higher order standing spin waves in buried dielectric ferromagnetic layers\bracite{deb2021b, zeus2022}. Previous experiments have reported a controllable attenuation and elongation of this hot electron pulse via additional metallic interlayers exhibiting strong electron-phonon coupling\bracite{fert2017, fert2019}. However, a thorough and simultaneous experimental assessment of the strain, heat and magnetization dynamics in such functional heterostructures is lacking. 

Here, we tailor the rapid distribution of the optically deposited energy within a Pt-Cu-Ni heterostructure by inserting an insulating MgO interlayer that stops the heat transport via electrons. We quantify the resulting strain response of the metallic layers via ultrafast X-ray diffraction (UXRD), to deduce the tailored heat transport and the shape of the strain pulses. The transient response of the buried functional Ni layer strongly depends on the position of the MgO interlayer either before or after the Cu layer. Utilizing the polar time-resolved magneto-optical Kerr effect (trMOKE), we probe the laser-induced change of the out-of-plane magnetization and identify the mechanisms (demagnetization, quasi-static strain and propagating strain), which drive coherent precession as function of the angle of the external magnetic field provided by a rotatable permanent magnet. We observe a pronounced dependence of the angle-dependent precession amplitude on the heterostructure design. We model the different responses of the heterostructures with a unique set of parameters by inputting the transient strain and heat - quantified by UXRD - into the Landau-Lifshitz Gilbert (LLG) equation. We discuss the concepts of how heterostructure design can control magnetization precession: The thickness of the propagation layer -- and hence the total multilayer stack -- sets the round trip time, which can be made resonant with the precession frequency that is tuned via the external magnetic field angle. The position of the MgO interlayer controls the thickness of the metal transducer and hence tailors the shape of the strain pulses, which are most efficient in driving the magnetization precession by matching the timing of both expansive and compressive strain to the phase of the precession. Generally the MgO interlayer \textcolor{black}{- or alternatively any other insulating material -} disables the ultrafast demagnetization and quasi-static expansion of the Ni layer as driving mechanisms that otherwise drive precession efficiently for out-of-plane external fields. 

Our report on the combined study of trMOKE and UXRD experiments is structured as follows: Section~\ref{sec:II} presents the magnetization dynamics triggered in three Pt-Cu-Ni heterostructures that only differ in the position of an insulating MgO layer. Section~\ref{sec:III} shows how UXRD quantifies the transient strain and heat in these heterostructures, before Section~\ref{sec:IV} discusses a model based on the LLG equation, where the extracted strain profile is used as an input parameter for the transient effective magnetic field. 
\begin{figure}[t!]
\centering
\includegraphics[width = \columnwidth]{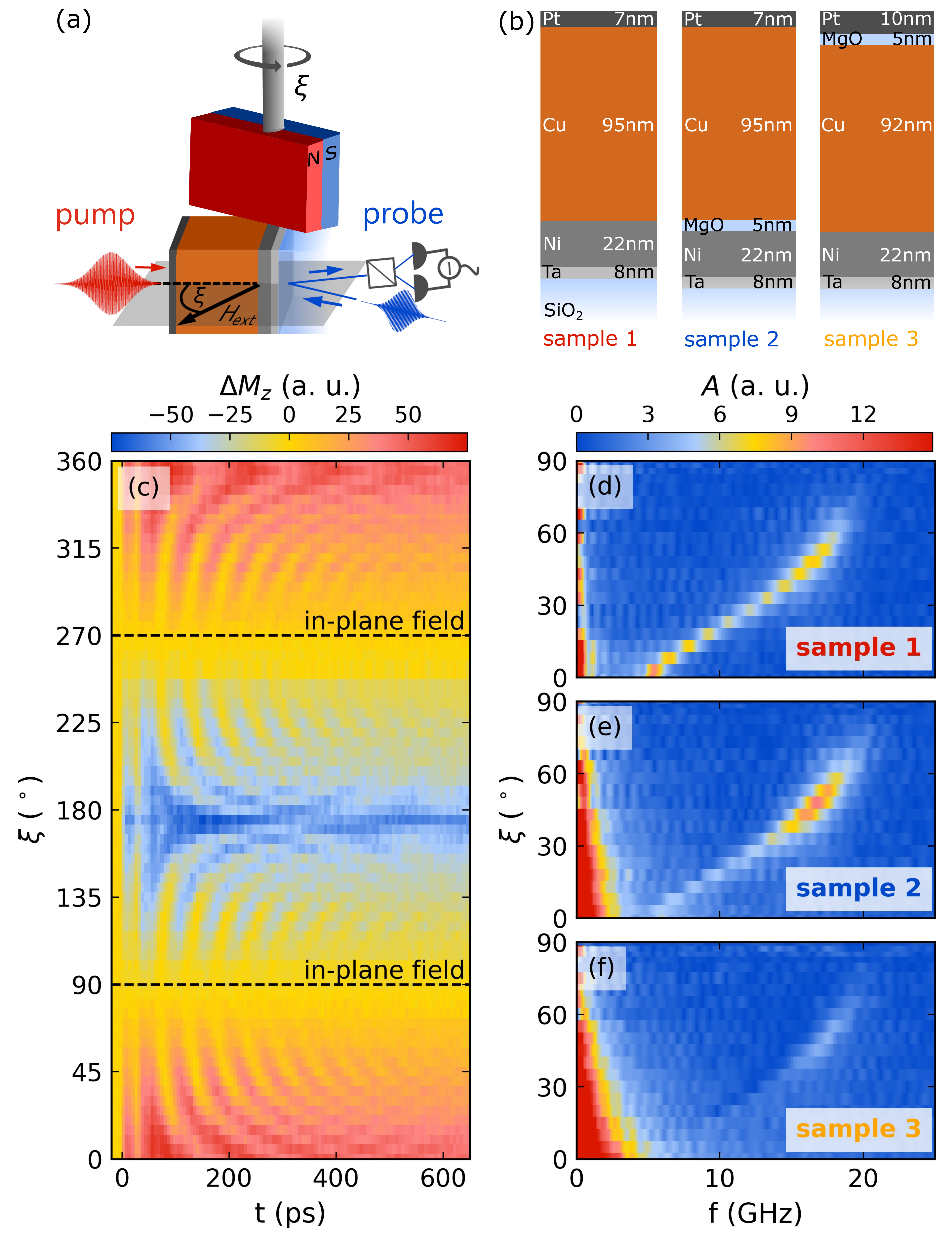}
\caption{\label{fig:fig_1_moke_experiment} Magnetization precession depending on the orientation of the external magnetic field: a) Sketch of the trMOKE experiment including a rotatable permanent magnet that provides a magnetic field of $\mu_\mathrm{0}H_\mathrm{ext} = 390\,\text{mT}$ at an angle $\xi$ with respect to the sample normal. b) Schematics of the three heterostructures with the Pt cap layer as the main light absorber and Cu as propagation layer for hot electrons. Sample 1 is a purely metallic structure and samples 2 and 3 contain an MgO layer to stop electron propagation.  c) Polar trMOKE measured for sample 1 for a full rotation of the external magnetic field by $360^\circ$. The oscillating out-of plane magnetization change $\Delta M_z(t)$ is color coded. See Figure~\ref{fig:fig_2_moke_meta}c) and d) for cuts at certain magnetic field angle $\xi$. d) Fourier-transformation of the trMOKE signal shows an increase of the precession frequency with in-plane orientation of the external field and a maximum FFT amplitude $A$ for $0^\circ$ and $47^\circ$. e,f) Same for samples 2 and 3, where only the maximum at $47^\circ$ is observed.}
\end{figure}

\section{\label{sec:II} Observation of the magnetization precession by trMOKE}
In the all-optical trMOKE experiment sketched in \textbf{Figure~\ref{fig:fig_1_moke_experiment}}a the three heterostructures shown in Figure~\ref{fig:fig_1_moke_experiment}b are excited by a near-infrared laser pulse with a pump energy density of $7.5\,\text{mJ}/\text{cm}^2$. The femtosecond pump pulse is mainly absorbed by the $7$ to $10\,\text{nm}$ thick Pt cap layer\bracite{pude2020}. While all heterostructures share the same structure of a Pt cap, a thick Cu transport and a $22\,\text{nm}$ buried Ni layer, they mainly differ by the ultrafast energy transfer between the Pt and Ni layer that is supported by the electrons in the Cu layer in sample 1 but suppressed by an insulating MgO interlayer either below or in front of the Cu layer in samples 2 and 3, respectively.

Figure~\ref{fig:fig_1_moke_experiment}c shows the polar trMOKE signal, i.e.\ the transient change of the out-of-plane magnetization $\Delta M_z$ in $\Delta\xi=5^\circ$ steps for a full rotation of the magnet for sample 1. The variation from out-of-plane ($0^\circ$ and $180^\circ$) to in-plane ($90^\circ$ and $270^\circ$) external magnetic fields shows a fourfold symmetry of the precession. Therefore, we only discuss the range from $0$ to $90^\circ$ in the following. The Fourier transform of these data in Figure~\ref{fig:fig_1_moke_experiment}d reveals an increasing precession frequency as the out-of-plane component of the external field decreases and the magnetization tilts towards the sample plane, which represents the magnetic easy axis. Since the external magnetic field $\mu_\mathrm{0}H_\mathrm{ext} = 390\,\text{mT}$ is below the saturation field $\mu_\mathrm{0}H_\mathrm{sat} = 620\,\text{mT}$, the magnetization is always oriented along an angle between the external field and the sample plane before laser excitation. Figure~\ref{fig:fig_1_moke_experiment}e,f show the same dependence of the precession frequency on the rotation angle $\xi$ for samples 2 and 3. The color code for the precession amplitude shows that the presence of the MgO layer in these samples suppresses the excitation of precession for an out-of plane orientation of the external field ($\xi=0^\circ$) which is pronounced for the purely metallic sample 1. All three samples, however, exhibit a maximum amplitude around $\xi=47^\circ$, with considerably smaller amplitude for sample 3.

\begin{figure}[t!]
\centering
\includegraphics[width = \columnwidth]{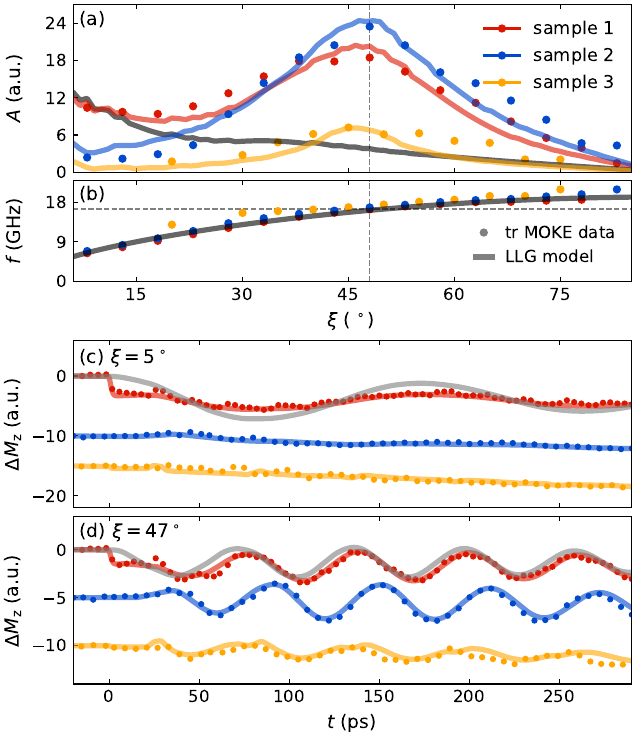}
\caption{\label{fig:fig_2_moke_meta} Heterostructure-dependent magnetization precession: a) Precession amplitude $A$ as function of the orientation of the external magnetic field. The extracted experimental data (symbols) are well reproduced via the LLG model described in Section~\ref{sec:IV} (solid lines) for all three heterostructures. All samples show a resonant enhancement at an angle, where the precession frequency (panel b) matches the inverse round-trip time of the propagating strain pulses, highlighted in Figure~\ref{fig:fig_3_uxrd}e) as black arrows. The black solid line denotes the hypothetical precession amplitude for sample 1 without any contribution from propagating strain pulses. b) Measured precession frequency $f$ (symbols) that is independent of the heterostructure, since all structures share the same Ni layer. c,d) Laser-induced change of the out-of-plane magnetization $\Delta M_z$ that displays the heterostructure-dependent demagnetization and magnetization precession at $\xi=5^\circ$ and $\xi=47^\circ$, respectively. The solid coloured lines represent the modeling described in Section~\ref{sec:IV}, while the grey solid lines represent the magnetization precession if we neglect the ultrafast demagnetization as driver of the precession in sample 1. The results are offset for clarity. }
\end{figure}

\textbf{Figure~\ref{fig:fig_2_moke_meta}} directly compares the driven magnetization precession in the three magnetic heterostructures: The extracted frequency in panel b confirms that the angle-dependence of the precession frequency is the same for all three Ni layers. Panel a provides a comparison of the precession amplitudes extracted from the FFT. Indeed all samples show a maximum at the external magnetic field angle $\xi=47^\circ$ for which the precession frequency matches the inverse round-trip time $T=L/v_\text{s}$ of longitudinal strain pulses through the multilayer stack (dashed horizontal line), which is essentially given by the total thickness $L$ and the thickness-weighted average sound velocity $v_\text{s}$. While in the presence of an MgO interlayer (samples 2 and 3) the precession amplitude rapidly decreases for larger and smaller $\xi$ values until it completely vanishes, we observe an additional increase of the precession amplitude towards $\xi=0^\circ$ in the absence of an MgO interlayer. These systematics already hint at the fact that the maximum at intermediate angles is a magneto-acoustic resonance that links the precession to the propagating strain pulses. In contrast, the low frequency precession close to $\xi=0^\circ$ is only enabled by the rapid heat transfer into Ni via hot electrons that leads to ultrafast demagnetization and an expansion.

The time-domain out-of-plane magnetization responses (symbols) shown in Figure~\ref{fig:fig_2_moke_meta}c,d provide further insights into the laser-induced driving of the magnetization precession. In the absence of an MgO interlayer, we clearly observe an ultrafast demagnetization originating from the rapid excitation of the Ni layer by hot electrons. This rapid excitation drives a precession at $\xi=5^\circ$, which is absent in samples 2 and 3, where electrons are stopped by the MgO interlayer and we only observe a slow demagnetization within hundreds of picoseconds (yellow and blue transients in Figure~\ref{fig:fig_2_moke_meta}). At $\xi=47^\circ$ we observe pronounced oscillations of varying amplitude for all three heterostructures. The precession amplitude even slightly increases from the first to the second maximum indicating the periodically acting driving force by propagating strain pulses. In addition, we note an immediate start of the precession upon laser-excitation in sample 1 and 2, while sample 3 clearly responds not earlier than at about $20\,\text{ps}$, which is the time it takes acoustic strain pulse to travel from the excited Pt cap layer through the Cu layer.

Similar qualitative assignments rationalizing the origin of precession are common in the literature. In the following, we quantify the laser-induced strain response of the Ni layer and its transient temperature increase by UXRD. Using the measured transient temperature rise and the laser-induced strain as input for our LLG model described in Section~\ref{sec:IV} that nicely captures the heterostructure-dependent precession (solid lines in Figure~\ref{fig:fig_2_moke_meta}) and unambiguously identifies the role of the different driving mechanisms.

\section{\label{sec:III} Characterization of strain and heat via UXRD}
In the UXRD experiment, the Pt-Cu-Ni heterostructures are excited by femtosecond p-polarized laser pulses with a central wavelength of $800\,\text{nm}$ as in the trMOKE experiment. We probe the transient layer-specific expansion by the shift of the layer-specific Bragg peaks with the reciprocal space slicing method\bracite{zeus2021} at a laser-driven table-top plasma X-ray source\bracite{schi2012}.

\begin{figure}[t!]
\centering
\includegraphics[width = \columnwidth]{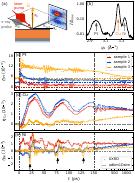}
\caption{\label{fig:fig_3_uxrd} UXRD quantifies heat transport and layer-specific strain: a) Sketch of the UXRD geometry: The detector records layer-specific Bragg peaks separated along the reciprocal space coordinate $q_z$. b) Integration of the detector signal along $q_y$ yields a reciprocal space slice, which encodes the mean out-of-plane lattice constant of the Pt, Cu and Ni layer within the heterostructures. c) Laser-induced strain response of the Pt layer (symbols) determined from the transient shift of the Bragg peak. The excitation fluence of $13.2\,\text{mJ}/\text{cm}^2$ for sample 1 and 2 was reduced to $7.5\,\text{mJ}/\text{cm}^2$ for sample 3 to avoid damage to the MgO interlayer. \textcolor{black}{The strain responses of the Pt capping layer are offset for clarity. } The solid lines represent our model of the transient strain response utilizing the modular \textsc{Python} library \textsc{udkm1Dsim}\bracite{schi2021} and a single set of thermo-elastic parameters collected in Table~\ref{tab:tab_1_sim_param}. \textcolor{black}{Introducing a Pt/MgO interface considerably slows down the cooling of Pt and the increased impedance mismatch enhances the initial fast oscillations.} d,e)  Laser-induced strain response for the Cu and Ni layers. Note the clear difference of the initial response of the Ni layer, which is expansive for sample 1, compressive for sample 2 and zero for sample 3. The black arrows indicate the arrival of the large amplitude strain pulse launched by the rapid Pt expansion. This periodicity of the Ni strain drives the magnetic resonance shown in Figure~\ref{fig:fig_2_moke_meta}a). \textcolor{black}{The transient strain responses of sample 1 and 2 are reproduced from our previous publication \cite{matt2023b}.}}
\end{figure}

\textbf{Figure~\ref{fig:fig_3_uxrd}}b displays the $(111)$ Bragg peaks of the Pt, Cu and Ni layers of sample 1. Their positions along the reciprocal space coordinate $q_z$ encode the average out-of-plane lattice constants $d$ of the respective layer via $q_z=2\pi/d$. The shift of the separated Bragg peaks along $q_z$ yields the layer-specific transient strain $\eta=\Delta d/d_0$ as relative change of the average out-of-plane lattice constant $\Delta d$ with respect to its value $d_0$ before excitation. Figure~\ref{fig:fig_3_uxrd}c--e displays the measured layer-specific strain responses (symbols) to an excitation of $13.2\,\text{mJ}/\text{cm}^2$ for sample 1 and 2 and $7.5\,\text{mJ}/\text{cm}^2$ for sample 3 choosen to avoid damage to the MgO interlayer. The solid lines represent the modeled transient strain response of the heterostructures utilizing the modular \textsc{Python} library \textsc{udkm1Dsim}\bracite{schi2021} and a single set of thermo-physical parameters stated in \textbf{Table~\ref{tab:tab_1_sim_param}} in section~\ref{sec:VI}.

In all three heterostructures, the optically deposited energy to the Pt layer induces an expansion that peaks at $1.5\,\text{ps}$ (Figure~\ref{fig:fig_3_uxrd}c) when the expansive strain pulses launched at the surface and the Pt-Cu interface have propagated through the Pt layer at the speed of sound $v_\text{Pt}=4.2\,\text{nm/ps}$\bracite{farl1966}. The following strain oscillation can be interpreted as a breathing of the Pt layer with a $T_\text{Pt}=2 L_\text{Pt}/v_\text{Pt} \approx 3\,\text{ps}$ period determined by the layer thickness $L_\text{Pt}$. The partial transmission of strain pulses through the impedance mismatched Pt-Cu or Pt-MgO interfaces leads to a decay of the oscillation of the Pt layer thickness within tens of picoseconds. This propagating strain pulse is superimposed on the quasi-static expansion of Pt due to heating. While the oscillatory Pt response is very similar for all three heterostructures, the quasi-static expansion is much increased for sample 3 where the MgO layer keeps the thermal energy within Pt, because it prohibits electronic heat transport from Pt into the thick Cu film.

This suppression of the rapid excitation of Cu as well as the buried Ni layer in sample 3 results in a delayed expansion within hundreds of picoseconds due to slow energy transfer into these layers via phonons. Therefore, the strain response within the first tens of picoseconds is dominated by the strain pulse with leading compression driven by the rapid expansion of the Pt layer. This strain pulse immediately compresses the adjacent Cu layer and subsequently also compresses the Ni layer when it enters at $20\,\text{ps}$ after having propagated through Cu. In sample 2, moving the MgO interlayer in front of Ni enables a rapid distribution of the optically deposited energy among Pt and Cu via hot electrons\bracite{pude2020}. This rapid excitation of Cu results in a rapid expansion (Figure~\ref{fig:fig_3_uxrd}d) that shortens the initial compression by the expanding Pt layer. Most importantly, the expansion of Cu drives a bipolar strain pulse that compresses the buried Ni layer already within the first picoseconds and superimposes with the strain pulse launched by Pt entering Ni at $20\,\text{ps}$. \textcolor{black}{The strong oscillation of the average strain of Cu with the period matching twice the transit time through the entire heterostructure is strongly suppressed in sample 1, where the MgO layer prevents energy transport to Cu via electrons.} 

Removing the MgO interlayer enables an additional fast diffusive transport of hot electrons from the optically excited Pt layer to the buried Ni through Cu as observed previously\bracite{pude2020}. Equilibration of the electronic system is very fast, and the large electronic heat capacity and the strong electron-phonon coupling in Ni originating from the large density of states of flat $3\text{d}$-bands close to the Fermi level rapidly localizes a large fraction of the deposited energy in Ni. \textcolor{black}{For our heterostructures with $100\,\text{nm}$ thick Cu layer, details such as initial ballistic or superdiffusive transport are not relevant, as we merely differentiate sub-picosecond heat transport by electrons from slower transport via phonons.} The large amount of energy deposited to Ni drives an expansion that launches an additional compression into the adjacent Cu layer extending its compression to $6\,\text{ps}$. In total, the strain response of Ni is the superposition of its quasi-static expansion due to heating and the strain pulses launched by Cu, Pt and Ni. The strain pulse emerging from Pt enters Ni at $20\,\text{ps}$ and is reflected at the substrate and the surface several times as indicated by the arrows in Figure~\ref{fig:fig_3_uxrd}e.

Our model of the layer-specific transient strain response (solid lines) reproduces these experimental observations for all three layers in all three heterostructures with a single parameter set. The model therefore yields the transient average electron $T^\text{el}_\text{Ni}(t)$ and phonon $T^\text{ph}_\text{Ni}(t)$ temperature as well as the average transient strain $\eta_\text{Ni(t)}$ of the buried Ni layer in all three heterostructures. In the following, these quantities validated by the good agreement of our model with the UXRD experiment serve as input for our LLG model of the heterostructure-dependent magnetization precession.

\section{\label{sec:IV} Modeling of the magnetization precession and Discussion}
In this section, we relate the heterostructure-dependent strain response and heating of the Ni layer quantified by UXRD in Section~\ref{sec:III} to the precession observed by polar trMOKE in Section~\ref{sec:II} and discuss how heterostructure design can control the driving mechanisms of magnetization precession. 

We model the precessional motion of the macrospin vector $\Vec{M}(t)=M_\text{sat} (m_x(t) \Vec{e}_x+m_y(t) \Vec{e}_y+m_z(t) \Vec{e}_z)$ that represents the average magnetization of the Ni layer in the heterostructures utilizing the LLG equation:
\begin{equation}
\frac{\partial \Vec{m}}{\partial t} = -\gamma \mu_0 \Vec{m}\times \Vec{H}_\text{eff} +\alpha \vec{m} \times \frac{\partial \vec{m}}{\partial t} \, ,
\label{eq:eq_1_LLG}
\end{equation}
where $\gamma$ represents the gyromagnetic ratio and $\alpha$ the phenomenological Gilbert damping. The effective field $\Vec{H}_\text{eff}(t)$ drives the magnetization precession and is determined by the material-specific free energy of the macroscopic magnetization $F\textsubscript{M}$ via $\mu_0 \Vec{H}_\text{eff}=-\Vec{\nabla}_M F\textsubscript{M}$. In our model of the macrospin in Ni, we consider a Zeeman energy originating from the external magnetic field $\Vec{H}_\text{ext}$, a shape anisotropy resulting from the thin film geometry and a magneto-acoustic contribution induced by the average out-of-plane strain within the Ni layer $\eta_\text{Ni}(t)$:
\begin{equation}
    \begin{split}
        \Vec{H}_\text{eff}(\Vec{m}, \eta) &= \Vec{H}\textsubscript{ext} - \Vec{H}\textsubscript{d}  - \Vec{H}\textsubscript{me} \\
        &= \Vec{H}\textsubscript{ext} - M\textsubscript{sat} m_z(t) \Vec{e}_z  - \frac{2 b_1\eta_\text{Ni}(t)}{\mu_0 M\textsubscript{sat}}\,m_z(t) \Vec{e}_z\; .
    \end{split}
\label{eq:eq_2_Heff}
\end{equation}
In agreement with previous studies in Ni\bracite{shin2022, shin2023} we neglect any magnetocrystalline anisotropy, which is supported by a negligible in-plane saturation field \textcolor{black}{(see Fig.~\ref{fig:fig_5_characterization}(b))} for all samples and a quantitative agreement of the out-of-plane saturation field of \textcolor{black}{$620\,\text{mT}$ (see Fig.~\ref{fig:fig_5_characterization}(a))} with $\Vec{H}\textsubscript{d}$ for the measured saturation magnetization $M_\text{sat}$ \textcolor{black}{(see Fig.~\ref{fig:fig_5_characterization}(b))}. The parameter $b_1$ determines the strength of the coupling of the laser-induced strain to the out-of-plane magnetization $M_z$. In addition to the magneto-elastic coupling, our model considers the demagnetization of Ni as driving mechanism of magnetization precession. The reduced length of the magnetization vector $\Vec{m}(T^\text{ph}_\text{Ni}(t))$ as function of the phonon temperature $T^\text{ph}_\text{Ni}$ introduces a change of the effective field via the changing demagnetization field.\footnote{Here we assume the static reduction of the magnetization $M(T)$ according to VSM measurements. We numerically tested that a very short-lived demagnetization that could be expected for ultrafast laser excitation would not change the slow precessional dynamics considerably.} The transient strain $\eta_\text{Ni}(t)$ and the transient temperature $T^\text{ph}_\text{Ni}(t)$ of the Ni layer are not adjustable parameters in our model: We use the time dependence that is quantified from modeling the laser-induced strain response of all three layers in the UXRD experiment. For all three heterostructures, we use the same set of parameters. \textcolor{black}{The experimental section~\ref{sec:VI} collects the measured magnetic properies of the 20 nm Ni film, which are essentially identical in all three heterostructures. From the UXRD experiment we have calibrated the spatio-temporal temperature $T(z,t)$ profile and in particular the average transient strain $\eta_\text{Ni}(t)$ of the Ni layer. When fitting the measured precession amplitude, this procedure calibrates the magneto-acoustic coupling parameter $b_1=-8.2\cdot10^6$\,J/m\textsuperscript{3}, in rough agreement with previous studies on polycrystalline Ni \cite{shin2023, vern2022}.}
\begin{figure}[t!]
\centering
\includegraphics[width = \columnwidth]{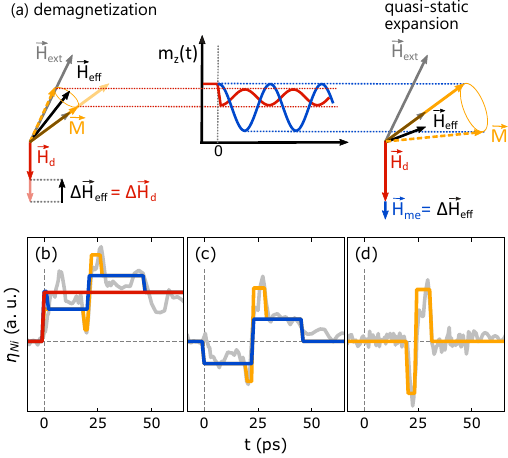}
\caption{\label{fig:fig_4_model} Driving mechanisms of the precession: a) Sketch of the change of the effective field induced by demagnetization (left) and quasi-static expansion (right). The dotted lines indicate the maximal excursions of the magnetization vector. The demagnetization yields a very rapid decrease on $m_z$, and a concomitant change $\Delta H_\text{eff}$, whereas the quasi-static expansion only induces a $\Delta H_\text{eff}$ with opposite sign according to Equation~\eqref{eq:eq_2_Heff}. The black and dark-grey arrows indicate $\Vec H_{eff}$ after and before switching on the demagnetization and quasistatic expansion, respectively. Note that strain pulses, i.e.\ time-dependent strain $\eta(t)$ yields similar time-dependent changes of the effective field with both polarities. b) The measured transient strain of the Ni layer in sample 1 is separated into a strong step-like quasi-static expansion (red), a long bipolar strain pulse launched by the thick Cu layer (blue) and a sharper strain pulse originating from Pt (yellow). c) Same for sample 2, where the quasi-static expansion is missing because MgO prevents energy flow into Ni. d) Same for sample 3, where only the strain pulse from Pt persists, as the MgO layer confines the energy there.}
\end{figure}

With an excellent match to the experimental data, Figure~\ref{fig:fig_2_moke_meta}a,b displays the modeled precession frequency and amplitude for all heterostructures as function of the angle $\xi$ of the external magnetic field. According to Equation~\eqref{eq:eq_2_Heff} the precession frequency increases with $\xi$ due to an increasing $H_\text{eff}$ as the demagnetization field $\Vec{H}_\text{d}$ only compensates the out-of-plane component of the external field while its in-plane component prevails. A comparison of the model to the time-domain out-of-plane magnetization response in Figure~\ref{fig:fig_2_moke_meta}c,d confirms also an excellent match regarding the phase of the precession and the temperature-induced demagnetization.

Based on our model of the  effective field  $H_\text{eff}$ in Equation~\eqref{eq:eq_2_Heff}, we distinguish two main contributions to the laser-induced change of the effective field: (i) the demagnetization of Ni originating from the laser-induced temperature-increase in Ni and (ii) the transient strain in Ni that is the superposition of a quasi-static expansion and propagating strain pulses. In the absence of an MgO interlayer, the rapid heating of Ni induces both demagnetization and quasi-static expansion. \textbf{Figure~\ref{fig:fig_4_model}}a illustrates their counteracting influence on the effective field in Ni. The ultrafast demagnetization shortens the magnetization vector $\Vec{m}$. This reduces the demagnetization field $\Vec{H}_\text{d}$ via the reduced out-of-plane component $m_z$. As a result, the effective field change $\Delta \Vec{H}_\text{eff}=\Delta \Vec{H}_\text{d}$ points along the positive $z$-direction and the shorter magnetization vector precesses around the new effective field $H_\text{eff,0}+\Delta \Vec{H}_\text{eff}$. In contrast, the magneto-elastic field $\Vec{H}\textsubscript{me}$ induced by the expansion of Ni causes an additional contribution to the effective field pointing along the negative $z$-direction. This opposite change of the effective field results in opposite starting directions of the precession corresponding to a phase shift of $\pi$. If we neglect the demagnetization as driving mechanism, the agreement of the model (grey solid line) with the experimentally observed precession in Figure~\ref{fig:fig_2_moke_meta}c,d is much worse. Especially, the precession amplitude at $\xi=5^\circ$ (Figure~\ref{fig:fig_2_moke_meta}c) is much larger without demagnetization, that counteracts the effective field due to quasi static strain. The fact that the precession for sample 1 and $\xi=5^\circ$ starts with an additional decrease of $m_z$ after the demagnetization indicates the dominance of the quasi-static strain over the demagnetization contribution. At $\xi=47^\circ$ (Figure~\ref{fig:fig_2_moke_meta}d) the model neglecting demagnetization leads to a severe phase shift. In Figure~\ref{fig:fig_2_moke_meta}a, the black solid line represents the modeled precession amplitude considering only the quasi-static expansion and demagnetization of Ni in sample 1. The maximum around $\xi=47^\circ$ is absent without propagating strain pulses, which provide an driving force of the magnetization precession that is resonant with the precession frequency at this angle. 

To rationalize the characteristic dependence of the precession amplitude on the heterostructure design, we reconsider the transient strain response of Ni $\eta_\text{Ni}(t)=\eta_\text{qs}(t)+\eta_\text{p}(t)$ quantified by UXRD (Section~\ref{sec:III}) that is the superposition of a quasi-static strain $\eta_\text{qs}$ and propagating strain pulses $\eta_\text{p}$ launched by rapidly expanding layers. Figures~\ref{fig:fig_4_model}b--d dissect the heterostructure-dependent $\eta_\text{Ni}(t)$ into contributions highlighted in different colors. While for sample 3 (panel d) only a sharp bipolar strain pulse launched by the expansion of the Pt layer is present, sample 2 (panel c) exhibits a superposition of this strain pulse from Pt and a pulse with longer period that is attributed to the expansion of the thicker Cu layer. For the all-metallic sample 1 (panel b), the ultrafast energy transfer to the Ni layer additionally yields its quasi-static expansion highlighted in red.

The UXRD of the Ni layer (Figure~\ref{fig:fig_3_uxrd}e) evidences a periodic contribution to the strain $\eta_\text{Ni}(t)$ in Ni highlighted by the arrows. The period is given by the round-trip time at the speed of sound through the metallic heterostructure of $57\,\text{ps}$ from the Ta-glass interface to the surface and back. This periodic driving force most efficiently drives the magnetization precession when it matches the precession frequency, which results in a maximum precession amplitude at $\xi=47^\circ$ and $f=17.5\,\text{GHz}$ for all three heterostructures. In an alternative picture, this maximum precession amplitude relates to considerable Fourier amplitude at the precession frequency from the repetition of the bipolar pulses with the same round trip time of about $57\,\text{ps}$ through the entire heterostructure \cite{kova2013,vern2022}. However, the efficiency of this magneto-acoustic resonance depends on shape of the strain pulse. The reduced precession amplitude for sample 3 in comparison to sample 2 highlights that despite its large amplitude the spatially narrow high frequency strain pulse launched by the Pt layer drives the magnetization precession less efficiently than the spatially extended strain pulse with low-amplitude, that is launched by the Cu layer. We attribute this to the the fact that the $\approx 40\,\text{ps}$ period of the Cu strain pulse efficiently matches the phase of the positive and negative excursions of the strain pulse to the phase of the magnetization precession. In contrast, within a small fraction of the precession period the large effects of the positive and negative strain in the bipolar pulse launched by the Pt layer nearly cancel.

%In the absence of an MgO interlayer (sample 1), the rapid energy deposition to Ni via hot electrons gives rise to a rapidly rising quasi-static expansion and to ultrafast demagnetization that both serve as additional driving mechanism of the magnetization precession. Simultaneously, the energy flow to Ni reduces excitation of Cu and hence attenuates the launched low frequency strain pulse, lowering its contribution to the transient effective field. This results in the reduced precession amplitude at $\xi=47^\circ$ with respect to sample 2 (cf.\ Figure~\ref{fig:fig_2_moke_meta}b). 

\section{\label{sec:V} Conclusion}
In summary, we conducted systematic UXRD and trMOKE experiments on a series of nanoscale metallic heterostructures with a buried functional magnetic layer, in order to achieve a consistent modeling of the magnetization dynamics based on the measured transient strain and temperature. The excellent agreement for varying heterostructure design and external magnetic field confirms a predictive power of the modeling. Here, we identified the mechanisms driving magnetization precession in Ni and demonstrated their control by tailoring the heat transport within Pt-Cu-Ni heterostructures via insulating MgO interlayers. 

The insulating MgO interlayer stops the heat transport via electrons, which disables the quasi-static expansion and ultrafast demagnetization as driving mechanisms by suppressing the rapid excitation of the buried Ni layer. In the absence of an MgO interlayer, the expansion and demagnetization provide counteracting mechanisms that drive the precession most efficiently for out-of-plane magnetic field orientations. The precession phase reveals the quasi-static expansion as the dominant driver in the explored fluence regime. This efficient magneto-acoustic mechanism is similarly active at an external field orientation that brings the precession into resonance with acoustic sound propagation through the multilayer stack. By considering the actual shape of the strain pulses quantified by UXRD, we moreover show that in addition to this resonance, magneto-acoustic efficiency is boosted roughly by a factor $3$ by phase matching, i.e.\ when positive and negative strain contributions each act on the magnetization vector for about half of the precession period.

Our experiment demonstrates the crucial role of laser-induced strain as driving mechanism of magnetization precession and highlights heterostructure design as a promising approach to tune the efficiency of driving the magnetization precession that is central to pave the way towards coherent magnetization switching via strain. In this publication, the heterostructure serving as effective tuneable transducer consists of simple metals and an insulating layer blocking the heat transport via electrons. In future applications, the controlled manipulation of buried layers by heterostructures can be extended. \textcolor{black}{Multi-pulse coherent-control scenarios can exploit the counteracting influence of demagnetization and strain\cite{matt2023c}.}
\textcolor{black}{A broader range of functional heterostructures can profit from our analysis when choosing materials with stress generation mechanisms or heat transport properties that can be tuned by external parameters such as temperature or electric and magnetic fields.}

\section{\label{sec:VI} Experimental Section}
\textit{Sample growth \textcolor{black}{and characterization}}: The metal heterostructures (Figure~\ref{fig:fig_1_moke_experiment}b) consisting of a $7$ to $10\,\text{nm}$ thick Pt cap layer, a $95\,\text{nm}$-thick Cu transport layer and a buried $22\,\text{nm}$ Ni layer were sputtered onto glass substrates (Corning 1737 AMLCD). The $8\,\text{nm}$ amorphous Ta seed layer ensures textured crystalline growth of Ni, Cu and Pt with (111)-orientation. 
\begin{figure}[h!]
\centering
\includegraphics[width = \columnwidth]{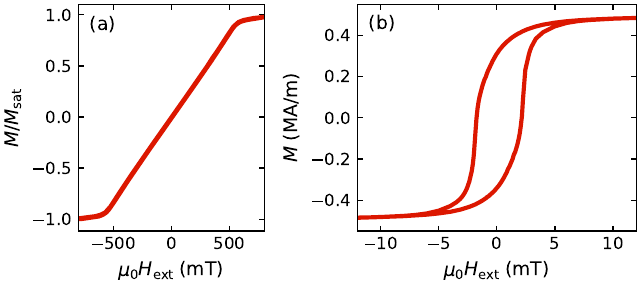}
\caption{\label{fig:fig_5_characterization} Magnetic characterization of the Ni layer (sample 1): (a) Out-of-plane hysteresis measured by polar MOKE and (b) in-plane hysteresis measured by SQUID vibrating sample magnetometry (VSM).}
\end{figure}

\textcolor{black}{As an input of the modelling of the magnetization precession, we characterized the magnetic properties of the Ni layers. Figure~\ref{fig:fig_5_characterization} exemplary displays the out-of-plane and in-plane hysteresis of sample 1 that are nearly identical for the other samples. We found a saturation magnetization of $M_\text{sat}=4.9\cdot10^5\,\text{A/m}$ via vibrating sample magnetometry (see Fig.~\ref{fig:fig_5_characterization}(b)). The shape anisotropy parametrized by the demagnetization field $\mu_0\Vec{H}_\text{d}= \mu_0 M_s m_z = 620\,\text{mT}$ matches the out-of-plane saturation field determined by polar MOKE (see Fig.~\ref{fig:fig_5_characterization}(a)). The in-plane hysteresis in Fig.~\ref{fig:fig_5_characterization}(b) indicate a negligible in-plane anisotropy. In addition, we performed ferromagnetic resonance (FMR) measurements on a very similarly grown Ni film without the Pt-Cu heterostructure determining the gyromagnetic ratio $\gamma/2\pi =30.8\,\text{GHz/T}$.}

\textit{MOKE Measurement}: In the all-optical trMOKE experiment sketched in Figure~\ref{fig:fig_1_moke_experiment}a,
the Pt-Cu-Ni heterostructures are excited from the Pt side by a $100\,\text{fs}$ pump-pulse with a central wavelength of $800\,\text{nm}$ at normal incidence with a repetition rate of $0.5\,\text{kHz}$, a beam footprint of $580 \times 480\,\text{\textmu m}^2$ and a pump energy density of $7.5\,\text{mJ}/\text{cm}^2$. We probe the transient out-of-plane magnetization of the Ni film through the glass substrate by detecting the polarization rotation of a reflected $400\,\text{nm}$ probe pulse. A commercial rotatable Neodymium permanent magnet (dimension: $60 \times 30 \times 15\,\text{mm}^3$) provides an external magnetic field of $\mu_\mathrm{0}H_\mathrm{ext} = 390\,\text{mT}$ at the probed sample region under an angle $\xi$ with respect to the surface normal.

\textit{UXRD Measurement}: In the UXRD experiment, the heterostructures are excited by an $60\,\text{fs}$ pump-pulse with a central wavelength of $800\,\text{nm}$ incident at $50^\circ$ with respect to the sample normal with a repetition rate of $1\,\text{kHz}$ and a beam footprint of $750 \times 840\,\text{\textmu m}^2$. The incident angle $\alpha_\text{in}$ of the $200\,\text{fs}$-long hard X-ray probe pulses with a photon energy of $\approx 8\,\text{keV}$ \textcolor{black}{provided by a table-top laser-based plasma x-ray source \cite{schi2012}} is kept fixed at $19.5^\circ$ to probe the strain response of the Pt layer and at $22.1^\circ$ for the Cu and Ni layers. At an intermediate angle we are able to record all three Bragg peaks simultaneously\bracite{matt2023b} on a pixelated area detector as displayed in the detector inset of Figure~\ref{fig:fig_3_uxrd}a that sketches the experimental geometry with the pump beam incident at $50^\circ$ with respect to the sample normal.

\textit{Modelling strain response}: To model the strain response, we use the modular \textsc{Python} library \textsc{udkm1Dsim}\bracite{schi2021} and a single set of thermo-physical parameters stated in Table~\ref{tab:tab_1_sim_param}. \textcolor{black}{In the first step, we calculate the absorption profile utilizing a transfer matrix approach \cite{khor2014} utilizing literature values for the refractive index $\Tilde{n}$ at $800\,\text{nm}$ given in Table~\ref{tab:tab_1_sim_param}. Starting from the calculated depth-dependent energy deposition,} we solve the diffusive two-temperature model\bracite{pude2020, matt2023b} to determine the spatio-temporal distribution of the laser-deposited energy within the heterostructures, which linearly relates to a spatio-temporal laser-induced stress via subsystem-specific Gr\"uneisen parameters $\Gamma_r$.\bracite{matt2023b} The solution of the elastic wave equation for these stress contributions yields the spatio-temporal strain response, which is translated into Bragg peak shifts via dynamical X-ray diffraction theory that yield the transient layer-specific strain as in the UXRD experiment. Details of our model approach were reported previously\bracite{matt2023b}.
\begin{table*}[t!]
\centering
\begin{ruledtabular}
\begin{tabular}{l c c c c c c}
 & Pt & Cu & Ni & Ta & MgO & glass \\
 \hline
$\Tilde{n}$ & 0.576+8.078i\bracite{wern2009} & 0.105+5.141i\bracite{wern2009} & 2.322+8.882i\bracite{wern2009} & 0.992+7.293i\bracite{wern2009} & 1.728\bracite{step1952} & 1.5\bracite{corn2002}\\
$\gamma\textsuperscript{S}$ (mJ\,cm\textsuperscript{-3}\,K\textsuperscript{-2})& 0.73\bracite{hohl2000} & 0.10\bracite{lin2008} & 1.06\bracite{hohl2000} & 0.38\bracite{bodr2013} & - & -\\
$C\textsubscript{ph}$ (J\,cm\textsuperscript{-3}\,K\textsuperscript{-1}) & 2.85\bracite{shay2016} & 3.44 & 3.94\bracite{mesc1981}  & 2.33\bracite{bodr2013} & 3.32 \bracite{barr1959} & 1.80 \bracite{corn2002}\\
$\kappa_\text{el}^0$ (W\,m\textsuperscript{-1}\,K\textsuperscript{-1}) & 66\bracite{dugg1970} & 396\bracite{hohl2000} &  81.4\bracite{hohl2000} & 52.0 & - & -\\
$\kappa\textsubscript{ph}$ (W\,m\textsuperscript{-1}\,K\textsuperscript{-1}) &5.0\bracite{dugg1970} & 5.0&  9.6\bracite{hohl2000} & 5.0 & 50 \bracite{slif1998}(2.5) & 1.0 \bracite{corn2002}\\
$g$ (PW\,m\textsuperscript{-3}\,K\textsuperscript{-1})& 375\bracite{zahn2021} & 95\bracite{lin2008} & 360\bracite{lin2008} & 100 & - &-\\
$\rho$ (g\,cm\textsuperscript{-3}) & 21.45 & 8.96 & 8.91 & 16.68 & 3.58 & 2.54 \bracite{corn2002}\\
$v\textsubscript{S}$ (nm\,ps\textsuperscript{-1})& 4.2\bracite{farl1966} & 5.2\bracite{over1955} & 6.3\bracite{neig1952} & 4.2\bracite{feat1963} & 9.1 \bracite{dura1936} & 5.7 \bracite{corn2002} \\
$\Gamma$\textsubscript{el} & 2.4\bracite{kris1979}(1.2) & 0.9\bracite{kris1979} & 1.4\bracite{wang2008} & 1.3\bracite{kris1979} & - & - \\
$\Gamma$\textsubscript{ph} & 2.6\bracite{nix1942} & 2.0\bracite{nix1941} & 1.8\bracite{wang2008} & 1.6\bracite{kris1979} & 1.7 \bracite{whit1966} & 0.3 \bracite{corn2002}\\
\end{tabular}
\end{ruledtabular}
\caption{Literature values for the physical parameters of the strain model: The complex refractive index $\Tilde{n}$, the Sommerfeld constant $\gamma\textsuperscript{S}$, the specific heat of the phonons $C\textsubscript{ph}$, the electron-phonon coupling constant $g$, and the electron $\kappa_e^0$ and phonon $\kappa\textsubscript{ph}$ heat conductivity determine the spatio-temporal energy distribution upon laser-excitation in the framework of a diffusive two-temperature model\bracite{pude2020,matt2023b}. The subsystem-specific Gr\"uneisen parameters $\Gamma$\textsubscript{el} and $\Gamma$\textsubscript{ph} linearly relate the spatio-temporal energy density to an elastic stress on the lattice driving a quasi-static expansion and strain pulses propagating with sound velocity $v\textsubscript{S}$ according to the elastic wave equation\bracite{matt2023b}. Values in brackets are optimized values for the strain modeling. The significant reduction of the thermal conductivity of the MgO interlayer is related to a Kapitza interface resistance.}
\label{tab:tab_1_sim_param}
\end{table*}

\begin{acknowledgments}
We acknowledge the DFG for financial support via No.\ BA 2281/11-1 and Project-No.\ 328545488 -- TRR 227, project A10. We thank Jamal Berakdar (project B06) for fostering the discussions about modeling the magnetization dynamics.
\end{acknowledgments}

\bibliography{references.bib}
\end{document}